\documentstyle[aps,prb,eqsecnum,epsf]{revtex}
\begin{document}
\title{Generalized Random-Phase Approximation Theory of Quasiparticle Spectral Functions: Application to 
Bilayer Quantum Hall Ferromagnets}
\author{Yogesh N. Joglekar$^{1,2}$, Allan H. MacDonald$^{1,2}$}
\address{$^1$ Department of Physics, \\ Indiana University, \\ Bloomington, IN 47405}
\address{$^2$ Department of Physics, \\ University of Texas at Austin, \\ Austin, TX 78705}
\maketitle
\begin{abstract}
We present a microscopic theory of ground-state spectral function of bilayer quantum Hall systems that 
includes interactions between Hartree-Fock quasiparticles and quantum fluctuations of the order parameter 
field. The collective modes in these systems are properly described only when fluctuations in direct 
and exchange particle-hole channels are taken into account. Using an auxiliary field functional integral 
approach, we present a generalization of the random phase approximation 
\emph{for quasiparticle self-energy} which captures fluctuations in both channels. We discuss its 
relationship to diagrammatic perturbation theory and an adiabatic approximation. We present simple 
analytical results for the quasiparticle self-energy and the renormalized order parameter that follow from 
this theory.
\end{abstract}


\section{Introduction}
\label{sec: intro}
The physics of a two-dimensional (2D) electron gas in a perpendicular magnetic field is unique in many 
respects. Bilayer electron systems consist of a pair of 2D electron gases
separated by a distance $d$ ($d\approx$ 100 \AA) which is comparable to the typical distance 
between electrons within one layer. In the presence of a strong magnetic field perpendicular to the 
layers, the kinetic energy of electrons is quenched and the physical properties of the system are 
determined by electron-electron interactions alone. Bilayer quantum Hall systems at Landau level 
filling factor $\nu=1$ exhibit a rich variety of broken symmetry states~\cite{kmoon,ky,nayak,kyang} and have
 been extensively studied over the past 
decade~\cite{kmoon,ky,nayak,kyang,qhereviews,haf,apb,brey,wenzee,ezawa,sqm,az,sgm,leon,ady,fogler}. 
For filling factor $\nu=1$, the ground state is fully spin polarized and the spin excitations are gapped 
because of Zeeman coupling of electron spins to the external magnetic field. The only remaining 
dynamical degrees of freedom are the discrete layer index and the intra-Landau level 
orbitals. We model the layer index using a pseudospin label where ``up'' denotes the symmetric bilayer 
state (S) and ``down'' denotes the antisymmetric bilayer state (AS). \emph{Note that this choice of 
quantization axis is different from the usual convention where ``up'' denotes a state localized in the top 
layer.} At {\it total} filling factor $\nu=1$, because of strong interlayer correlations, the ground state 
of the system exhibits \emph{spontaneous} interlayer phase coherence (easy-plane ferromagnetism in 
the pseudospin language) and is incompressible~\cite{apb,sqm}. At a critical layer separation 
$d_{cr}$ the system undergoes a phase transition from this incompressible quantum Hall state with 
pseudospin ferromagnetism to a disordered compressible state~\cite{apb,sqm}, possibly with 
other more exotic intervening states~\cite{nayak}.
	
In this paper we present a theory which describes effect of quantum fluctuations on the Hartree-Fock 
quasiparticles of the $\nu=1$ bilayer. Some of the results presented here were published 
earlier~\cite{ynjahm}. Our theory is based on an approximate expression for the quasiparticle self-energy 
which includes the effect of quantum fluctuations and generalizes the random phase approximation (RPA) 
to cases when both direct and exchange fluctuations are important. This self-energy modifies the 
spectral function of the mean-field quasiparticles, which in turn renormalizes the physical properties 
of the system. The most important macroscopic parameter characterizing the phase-coherent state is the 
pseudospin polarization. We use pseudospin polarization normalized to its mean-field value as a 
dimensionless order parameter $m_z$. This order parameter is related to quasiparticle spectral weights 
integrated up to the Fermi energy, $m_z=(n_S-n_{AS})$, where $n_\sigma$ denote occupation numbers for 
the symmetric and antisymmetric states. Therefore the order parameter is sensitive 
to changes in the quasiparticle spectral function. In the mean-field approximation the quasiparticles 
have sharply defined energies and the spectral functions are $\delta$-functions at these energies. In the 
mean-field ground state, all electrons occupy the symmetric state in the lowest Landau 
level, which corresponds to all pseudospins pointing along the positive $z$-axis, $m_z=1$. We will find 
that when we include the effect of collective fluctuations, the spectral functions develop a continuum 
piece and the spectral weight at the $\delta$-function peak is reduced. We will calculate the renormalized 
order parameter by evaluating the fluctuation correction to the mean-field spectral function. 

The generalized random phase approximation for quasiparticle self-energy presented in this paper can be 
usefully adopted to study the effect of quantum fluctuations on mean-field quasiparticles whenever the 
collective-mode behavior is determined by fluctuations in more than one particle-hole channels; for 
example, to analyze the effect of fluctuations on Wigner crystal mean-field quasiparticles at low 
filling factors. Traditionally random phase approximations have been used either to take into account the 
screening of Coulomb interaction~\cite{mahan} where only direct-channel interactions determine 
the physics, or the effect of spin-waves on a ferromagnetic mean-field state~\cite{he,kasner} where the 
low-energy physics is solely determined by exchange interactions. The generalization which 
we present here systematically takes into account fluctuations in both channels. We study the bilayer 
quantum Hall system as a test case because its translation invariance combined with the absence of dispersion 
in the Landau level bands permits many elements of this calculation to be performed analytically, and 
because the effect of quantum fluctuations on the mean-field ground state can be tuned over a large 
range simply by changing the layer separation. In the $d=0$ limit, since the interaction Hamiltonian is 
pseudospin invariant, the mean-field ground state is exact. At finite $d$, because of the 
difference between the intralayer Coulomb interaction $V_A(\vec{q})=2\pi e^2/q$ and the interlayer 
Coulomb interaction $V_E(\vec{q})=V_A(\vec{q})e^{-qd}$, the symmetry of the interaction Hamiltonian is 
reduced from SU(2) to U(1), and the order parameter $m_z$ does not commute with the 
interaction. Therefore the pseudospin-polarized mean-field state is not an eigenstate of the Hamiltonian 
and in particular, it is not the ground state. The exact ground state of this system incorporates quantum 
fluctuations in the quantum Hall ferromagnet's order parameter field. 

The paper is organized as follows. In section~\ref{sec: affia} we describe the functional integral 
approach to this problem. It is known that the collective modes in these systems are properly 
described only when both electrostatic and exchange fluctuations are taken into account~\cite{haf,apb}.
 No standard Hubbard Stratonovich (HS) transformation can treat both effects 
simultaneously~\cite{negele}. Section~\ref{sec: affia} also summarizes the generalized Hubbard-Stratonovich 
which we employ and establishes some of the notation which we will use. In this section we present 
 the generalized random phase approximation (GRPA) for the quasiparticle self-energy, which takes into 
account fluctuations in both channels and is based on an approach developed previously by Kerman 
{\it et al.}~\cite{kerman} Section~\ref{sec: grpa} summarizes the relationship of this GRPA with
 diagrammatic perturbation theory. In section~\ref{sec: adls} we apply our formalism to a bilayer 
system at filling factor $\nu=1$. We evaluate the effect of fluctuations on the quasiparticle spectral 
functions and renormalization of the order parameter due to these fluctuations. Section~\ref{sec: aa} 
summarizes an adiabatic approximation for the quasiparticle self-energy in a quantum Hall ferromagnet and 
its close relationship with the two more general approaches we discuss. We conclude the paper with a brief 
discussion in section~\ref{sec: discussion}.


\section{Auxiliary Field Functional Integral Approach}
\label{sec: affia}
In bilayer systems, the anisotropy of the Coulomb interaction in pseudospin space makes it necessary 
to treat fluctuations in the Hartree and exchange channels on an equal footing. A general way to 
treat fluctuations around an ordered mean-field is the Hubbard Stratonovich 
transformation~\cite{negele}. In this approach, by introducing an integral over a Bose field $\phi$,
 we convert the two-body fermionic interaction term into a one-body term coupled to this field. Since 
the effective Hamiltonian obtained is quadratic in fermion operators, the trace over these degrees 
of freedom can be performed exactly and we obtain an effective action for the auxiliary Bose field. 
Stationary phase approximations to the bosonic action give various mean-field theories. The nature of 
the mean-field state depends upon the way the Hubbard Stratonovich transformation is implemented and the 
predictions for collective excitations depend qualitatively on the mean-field around which Gaussian 
fluctuations in the Bose field are considered. For example, fluctuations in the exchange mean-field 
produce spin-wave excitations in ferromagnets, whereas fluctuations in the Hartree mean-field lead to 
plasmons in an electron gas. For a bilayer system, we need a method which treats both Hartree and exchange 
fluctuations on an equal footing. It is known that the standard HS transformation does \emph{not} 
yield a Hartree-Fock mean-field and cannot capture \emph{both} Hartree and the exchange 
fluctuations~\cite{negele}.

Kerman, Levit and Troudet have presented a generalization of the HS transformation which overcomes these
limitations~\cite{kerman}. To establish our notation~\cite{ya} 
(which differs from that of Kerman {\it et al}.) we briefly review the previous work which describes 
fluctuation corrections to mean-field approximations for the grand potential. Our approach, on the other 
hand, enables us to systematically improve upon mean-field approximations for correlation functions.

Consider a many fermion Hamiltonian in second quantized form
\begin{equation}
\label{eq: affia1}
\hat{H}= \sum_{\alpha\beta}K_{\alpha\beta}c^{\dagger}_\alpha c_\beta +
\frac{1}{2}\sum_{\alpha\beta\gamma\delta}\langle\alpha\beta|V|\gamma\delta\rangle 
c^{\dagger}_\alpha c^{\dagger}_\beta c_\delta c_\gamma,
\end{equation}
where $K$ is a one-body (kinetic) term and $V$ is a two-body interaction. Here $\alpha$ denotes 
(a set of) single-particle quantum numbers. For example, in the case of a bilayer system 
$\alpha=(n,k,\sigma)$ where $n$ is the Landau level index, $k$ is the intra-Landau level index and 
$\sigma$ is the pseudospin label. Introducing pair-labels $a=(\alpha'\alpha)$ and the density matrix 
$\hat{\rho}_a=c^{\dagger}_{\alpha'}c_{\alpha}$, the kinetic term becomes $K_a\hat{\rho}_a$ where 
summation over repeated index is assumed. Similarly the interaction term can be written as 
$V_{ab}\hat{\Omega}_{ab}$ where $\hat{\Omega}_{ab}=:\hat{\rho}_a\hat{\rho}_b:$ is the normal ordered 
two-body term and $V_{ab}=\langle\alpha'\beta'|V|\alpha\beta\rangle$. Then in an obvious matrix 
notation, we write (~\ref{eq: affia1}) as $\hat{H}=K\hat{\rho}+V\hat{\Omega}/2$. The partition 
function in the grand canonical ensemble is given by
\begin{equation}
\label{eq: affia2}
Z=\mbox{Tr }e^{\alpha\hat{N}-\beta\hat{H}}=\lim_{\epsilon\rightarrow 0}
\mbox{Tr }e^{\alpha\hat{N}}T\prod_{j=1}^{M}
\left[1-\epsilon K\hat{\rho}_j-\frac{\epsilon}{2}V\hat{\Omega}_j\right],
\end{equation}
where $T$ denotes the imaginary-time ordered product, $\hat{N}$ is the number operator, 
$\epsilon=\beta/M$ and in the thermodynamic limit $\alpha/\beta\rightarrow\mu$, the chemical potential. 
Since the number operator commutes with the Hamiltonian, we can expand only the $e^{-\beta\hat{H}}$ 
term. The central idea of this generalized Hubbard transformation is to expand around an 
\emph{arbitrary} two-body interaction $U$, thus treating $K\hat{\rho}+U\hat{\Omega}/2$ as the dominant 
term and $(V-U)\hat{\Omega}/2$ as a perturbation. Using the Gaussian identity
\begin{equation}
\label{eq: affia3}
1 =\frac{1}{\sqrt{\det{U}}}\int\prod_{\gamma\delta}
\frac{d\phi_{\gamma\delta}(j)}{\sqrt{2\pi/\epsilon}}
\exp\left[-\frac{\epsilon}{2}\sum_{\alpha\beta\alpha'\beta'}
\phi_{\alpha\beta}(j)U^{-1}_{\alpha\alpha'\beta\beta'}\phi_{\alpha'\beta'}(j)\right]
\end{equation}
at each time index $j$, a typical term in Eq.(~\ref{eq: affia2}) becomes
\begin{equation}
\label{eq: affia4}
\left[1-\epsilon K\hat{\rho}_j-\frac{\epsilon}{2}V\hat{\Omega}_j\right] 
= 
\int{\cal D}\phi_j e^{-\epsilon\phi_j U^{-1}\phi_j/2}
\left[
1-\epsilon K\hat{\rho}_j +\epsilon\phi_j\hat{\rho}_j-\frac{\epsilon^2}{2}V\hat{\Omega}_j
 \frac{\phi_j U^{-1}\phi_j}{{\cal N}^2}\right].
\end{equation}
Here ${\cal N}^2={\cal N}\times{\cal N}$ stands for a repeat sum over the ${\cal N}$ 
single-particle state labels in the Hilbert space and ${\cal D}\phi_j$ implies both a product over 
single-particle labels and the relevant normalization factors. Introducing these fields at each 
time-step $j$ we get 
\begin{eqnarray}
\label{eq: affia5}
Z & = & \lim_{\epsilon\rightarrow 0}\int\prod_{j=1}^{M}{\cal D}\phi_j 
\exp\left(-{\cal S}[\phi]\right),  \\
\label{eq: affia6}
{\cal S}[\phi] & = & \frac{\epsilon}{2}\sum_{j=1}^{M}\phi_j U^{-1}\phi_j 
-\ln\mbox{Tr }e^{\alpha\hat{N}} T \prod_{j=1}^{M} 
\left[1-\epsilon K\hat{\rho}_j+\epsilon\phi_j\hat{\rho}_j-\frac{\epsilon^2}{2}V\hat{\Omega}_j 
\frac{\phi_j U^{-1}\phi_j}{{\cal N}^2}\right].
\end{eqnarray}
We stress that since $\langle\phi^2\rangle\approx 1/\epsilon$, the limit $\epsilon\rightarrow 0$ 
in Eq.(~\ref{eq: affia5}) must be taken \emph{after} all $\phi$ integrals are done.

Systematic approximations to the grand potential are obtained by expanding the bosonic action 
(~\ref{eq: affia6}) around its minimum. In the limit as $\epsilon\rightarrow 0$, the configuration of 
fields $\phi^{0}$ which minimizes the action ($\partial{\cal S}/\partial\phi|_{\phi^{0}} = 0$) is 
given by
\begin{equation}
\label{eq: affia7}
\phi^{0}_{\alpha\beta}(j)=U_{\alpha\alpha'\beta\beta'}
\langle\hat{\rho}_{\alpha'\beta'}(j)\rangle_{\phi^{0}}, \hspace{2cm} j=1,\ldots,M ;
\end{equation}
where $\langle\rangle_{\phi^{0}}$ indicates thermal average with the mean-field Hamiltonian 
$\hat{h}_i=(K-\phi^{0}_i)\hat{\rho}$. We emphasize that Eq.(~\ref{eq: affia4}) is an exact identity; 
since the variance of the auxiliary field is proportional to $1/\epsilon$, the interaction term with 
prefactor $\epsilon^2$ is of the same order as the kinetic term with the 
prefactor $\epsilon$ after all the $\phi$-integrals are done. The stationary phase approximation, however, 
corresponds to replacing the measure ${\cal D}\phi_j$ by ${\cal D}\phi_j\delta(\phi_j-\phi^{0}_j)$. 
Therefore, in the stationary phase approximation, interaction term with the prefactor $\epsilon^2$ can be 
neglected and does not contribute to the mean-field solution. It is clear from Eqs.(~\ref{eq: affia7}) 
that the mean-field Hamiltonian is solely determined by the trial interaction $U$ and is independent 
of the actual two-body interaction $V$. We get the Hartree (exchange) mean-field by choosing 
$U=-V$ ($U=+V^{ex}$), whereas $-U=(V-V^{ex})=V^{A}$ (the antisymmetrized interaction) gives the 
Hartree-Fock mean-field. 

We improve upon the stationary phase approximation by considering quadratic fluctuations around the 
mean field $\phi^{0}$. These fluctuations, of course, depend upon the true microscopic interaction $V$. 
Expanding the action (~\ref{eq: affia6}) to second order in the fluctuating fields 
$\phi_j=\phi^{0}+\delta\phi_j$ gives 
\begin{equation}
\label{eq: affia8}
\frac{\partial^2{\cal S}}{\partial\phi_j\partial\phi_{j'}}\equiv\epsilon M(j,j')=
\epsilon\left[
\delta_{jj'}U^{-1}-\epsilon\left(1-\delta_{jj'}\right)D^{kk'}+\epsilon\delta_{jj'}S
\right],
\end{equation}
where only time indices are explicitly shown. The matrices $D$ and $S$ are defined by
\begin{equation}
\label{eq: affia9}
D^{jj'}_{\alpha\beta\gamma\delta}=
\langle\hat{\rho}_{\alpha\beta}(j)\hat{\rho}_{\gamma\delta}(j')\rangle_{\phi^0} - 
\langle\hat{\rho}_{\alpha\beta}(j)\rangle_{\phi^0}
\langle\hat{\rho}_{\gamma\delta}(j')\rangle_{\phi^0},
\end{equation}
and
\begin{equation}
\label{eq: affia10}
S_{\alpha\beta\gamma\delta}=
U^{-1}_{\alpha\beta\gamma\delta}\frac{\langle V\hat{\Omega}\rangle_{\phi^0}}
{{\cal N}^2} + 
\langle\hat{\rho}_{\alpha\beta}(j)\rangle_{\phi^0}
\langle\hat{\rho}_{\gamma\delta}(j')\rangle_{\phi^0},
\end{equation}
where $\alpha$,$\beta$ are the single-particle labels and $j,j'$ stand for the time indices. The 
particle-hole response function $D$ is the same as the Hartree-Fock mean-field susceptibility and the 
fluctuation matrix $M^{-1}$ is essentially the propagator for collective modes around the mean-field. 
It is clear from Eqs.(~\ref{eq: affia8}), (~\ref{eq: affia9}) and (~\ref{eq: affia10}) that properties 
of the collective modes are determined by the mean-field $\phi^{0}$ and the true microscopic interaction 
$V$. We note that among all trial potentials, only the antisymmetrized interaction optimizes the quasiparticle 
grand potential, $\Omega_{qp}(U)={\cal S}[\phi^{0}]+ \mbox{tr}S/2$, obtained from the action 
(~\ref{eq: affia6}). Therefore, we will concentrate on the Hartree-Fock mean field which corresponds to 
$U=-V^{A}$. 

We now extend this approach to the evaluation of the one-particle Green's function 
${\cal G}_{ab}(\tau_a,0)=+\langle Tc_a(\tau_a)c^{\dagger}_b(0)\rangle$. Starting with this definition 
and introducing auxiliary fields at each time step, we get
\begin{eqnarray}
\label{eq: affia11}
{\cal G}_{ab}(\tau_a) &= &
\lim_{\epsilon\rightarrow 0}\frac{1}{Z}\mbox{Tr } T\left[\prod_{j=1}^{M}
\left(1-\epsilon\hat{H}_j+\epsilon\frac{\alpha}{\beta}\hat{N}_j\right)
c_a(\tau_a)c^{\dagger}_b(0)\right], \\
\label{eq: affia12}
& = & \lim_{\epsilon\rightarrow 0}\frac{1}{Z}\int{\cal D}\phi e^{-{\cal S}[\phi]}
\langle\langle c_a(\tau_a)c^{\dagger}_b(0)\rangle\rangle_\phi,
\end{eqnarray}
where ${\cal S}$ is the bosonic action (~\ref{eq: affia6}) and 
$\langle\langle c_a(\tau_a)c^{\dagger}_b(0)\rangle\rangle_\phi$ represents the finite-$\epsilon$ 
expression for thermal average with the auxiliary Hamiltonian $\hat{h}_j=(K-\phi^{0}_j)\hat{\rho}$. 

\begin{figure}[h]
\begin{center}
\epsfxsize=3in
\hspace{0.1cm}
\epsffile{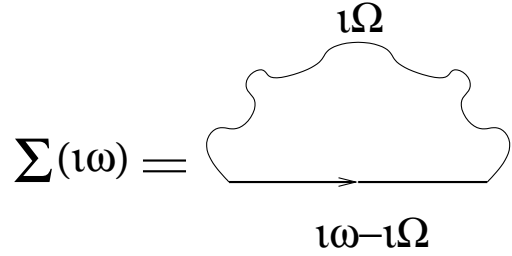}
\vspace{-2cm}
\caption{Schematic representation of self-energy approximation (~\ref{eq: affia14}). The directed solid  
line represents the mean-field Green's function ${\cal G}^{0}(i\omega-i\Omega)$ and the wavy line 
represents the collective-mode propagator $M^{-1}(i\Omega)$. The matrix elements at the vertex do no 
depend upon the energy transfer $\Omega$.}
\label{fig: qpsw}
\end{center}
\end{figure}

Systematic approximations to the exact Green's function are obtained by expanding the action ${\cal S}$ and 
the $\phi$-dependent thermal average around the mean-field value $\phi^{0}$. At the mean-field level, 
we obtain the Green's function of non-interacting quasiparticles in the self-consistent field $\phi^{0}$
\begin{equation}
\label{eq: affia13}
{\cal G}^{0}_{ab}(i\omega_n)=(-i\omega_n+K-\phi^{0}-\mu)^{-1}_{ab},
\end{equation}
where $\omega_n=\pi(2n+1)/\beta$ is a fermionic Matsubara frequency. 
To consider the effect of fluctuations, we expand the action (~\ref{eq: affia6}) to second order in 
fluctuating fields $\phi_j=\phi^{0}+\delta\phi_j$, replace the mean-field Green's function by a 
$\phi$-dependent Green's function
\begin{equation}
\label{eq: affia14}
{\cal G}_{ab}(i\omega_n) = \left(-i\omega_n+K-\phi^{0}-\delta\phi-\mu\right)^{-1}_{ab}
  = \left({\cal G}_{0}^{-1}-\delta\phi\right)^{-1}_{ab},
\end{equation}
and expand (\ref{eq: affia14}) in powers of $\delta\phi$ to \emph{all} orders. The quadratic expansion 
of the action (~\ref{eq: affia6}) gives non-interacting collective modes around the mean-field 
$\phi^{0}$. Expanding the $\phi$-dependent Green's function gives an effective interaction between the 
mean-field quasiparticles and the collective modes. The resulting integrals over the fluctuating fields are 
performed using Wick's theorem. In this approximation the Green's function has a self-energy that is the 
sum of all 1-particle irreducible diagrams obtained from mean-field Green's function ${\cal G}^{0}$ and 
the collective-mode propagator $M^{-1}$. We approximate this self-energy by the first irreducible diagram, 
which describes a single scattering of a collective mode and a mean-field quasiparticle 
(Figure~\ref{fig: qpsw}). The contribution from such a diagram is given by
\begin{equation}
\label{eq: affia15}
\Sigma_{ab}(i\omega_n)=-\frac{1}{\beta}\sum_{i\Omega_n}\sum_{\alpha'\beta'}
{\cal G}_{\alpha'\beta'}^{0}(i\omega_n-i\Omega_n)M^{-1}_{a\alpha',b\beta'}(i\Omega_n) 
\end{equation}
where $\Omega_n=2\pi n/\beta$ is a bosonic Matsubara frequency. 
Eq.(~\ref{eq: affia15}) is a general expression for self-energy of a quasiparticle because of its 
interactions with the collective modes around \emph{any} mean-field. We emphasize once again that this 
calculation is to be done with a finite value of $\epsilon$ and that the limit $\epsilon\rightarrow 0$ is 
taken only after all the $\phi$-integrals are evaluated. Since the dimension of the matrix $M$ is of order 
$1/\epsilon$ 
and since the matrix elements of $M$ are dependent on $\epsilon$, we have to be careful about elements of the 
inverse matrix. As we are interested in fluctuations around the Hartree-Fock mean field, we use $U=-V^{A}$ and 
Eq.(~\ref{eq: affia8}) for the collective-mode propagator, and arrive at the following expression for the 
quasiparticle self-energy
\begin{equation}
\label{eq: affia16}
\Sigma_{ab}(i\omega_n)=\frac{1}{\beta}\sum_{i\Omega_n}\sum_{\alpha'\beta'}
{\cal G}_{\alpha'\beta'}^{0}(i\omega_n-i\Omega_n)\left[\left(1+V^{A}D\right)^{-1}V^{A}-V^{A}\right]_
{a\alpha',b\beta'}(i\Omega_n).
\end{equation}
Here we have used the identity $A^{-1}_{ij}=\partial\ln\det A/\partial A_{ji}$, and the fact that 
$\ln\det M=\ln\det[1+V^{A}D]-\mbox{tr}(V^{A}D)$. We will call (~\ref{eq: affia16}) as the 
\emph{generalized random phase approximation} (GRPA) for quasiparticle self-energy. It is clear from 
Eq.(~\ref{eq: affia16}) that the GRPA self-energy includes diagrams with $n\geq 2$ interaction terms, and 
in particular, it does \emph{not} contain the mean-field self-energy contribution. This approximation for the 
self-energy arises naturally in our auxiliary field functional integral approach. In the next section we 
discuss the corresponding approximation in diagrammatic perturbation theory.


\section{Generalized Random Phase Approximation}
\label{sec: grpa}
In this section we describe the generalized random phase approximation for the quasiparticle self-energy in 
diagrammatics language and relate it to expression (~\ref{eq: affia16}) derived in the preceding 
section. In the case of particle-hole response functions~\cite{haf} or the grand potential~\cite{ya}, 
appropriate conserving generalizations of random phase approximation which take into account fluctuations 
in both, direct and exchange, particle-hole channels have been discussed in the literature. 

Similar generalizations for the quasiparticle self-energy are less than transparent. Several approaches 
and approximations have been used in the past. Some start from the ``bubble'' RPA which describes the 
screening and use vertex-corrected bubbles, while others start from the ``ladders'' RPA which 
captures the spin-wave dynamics and include the Hartree corrections by using screened interaction. We are 
not aware of a systematic generalization for the self-energy. 

\begin{figure}[h]
\begin{center}
\epsfxsize=3in
\hspace{1cm}
\epsffile{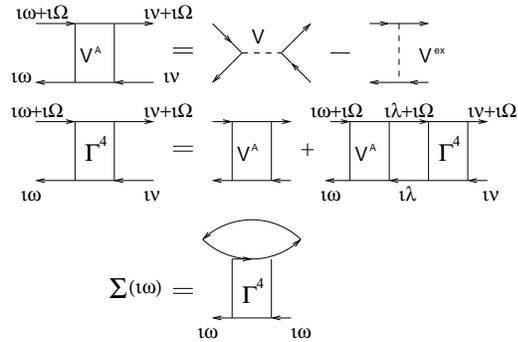}
\vspace{-2cm}
\caption{Diagrammatic summary of the GRPA for self-energy. We use the antisymmetrized interaction as the 
bare scattering vertex, $\Gamma^{(4)}_0=V^{A}$, to take into account fluctuations in the direct and the 
exchange channels. The self-energy is obtained from the scattering vertex $\Gamma^{4}$ by contracting 
the incoming and outgoing labels on the top.}
\label{fig: grpa}
\end{center}
\end{figure}

The appropriate generalization of the RPA self-energy which we propose is summarized diagrammatically in
 Fig.~\ref{fig: grpa}. The physical content of this self-energy is determined solely by the nature of 
fluctuations encoded in the four-point vertex $\Gamma^{(4)}(\vec{q},i\Omega_n)$. We use the antisymmetrized 
interaction as the bare four-point vertex, $\Gamma^{(4)}_0=V^{A}$. The self-energy obtained from the bare 
four-point vertex by contracting the incoming and outgoing labels on the top is the same as the the 
Hartree-Fock mean-field approximation for the self-energy. If instead of the antisymmetrized interaction, 
we use the direct (exchange) interaction as the bare vertex, we will get the Hartree (exchange) mean-field 
self-energy. We encode the fluctuations in direct and exchange channels into the four-point vertex by 
summing all the particle-hole ladders with the antisymmetrized interaction. In a matrix notation, the 
equation for the four-point vertex is 
\begin{equation}
\label{eq: grpa1}
\Gamma^{(4)} = V^{A} - V^{A}\cdot D\cdot \Gamma^{(4)}
\end{equation}
where $D$ is the Hartree-Fock mean-field susceptibility and we have suppressed all single-particle and 
imaginary-time indices. Eq.(~\ref{eq: grpa1}) can be formally solved 
\begin{equation}
\label{eq: grpa1.5}
\Gamma^{(4)}=\left(1+V^{A}\cdot D\right)^{-1} V^{A}. 
\end{equation}
In general, the matrix $(1+V^AD)$ is not diagonal its indices and therefore it is not possible to obtain an 
analytic expression for the four-point vertex. The GRPA self-energy is obtained from the four-point 
vertex by contracting its incoming and outgoing lines on the top and subtracting the mean-field 
self-energy contribution from the resulting self-energy. The formal expression for fluctuation self-energy 
is given by 
\begin{equation}
\label{eq: grpa2}
\Sigma_{ab}(i\omega_n)=\frac{1}{\beta}\sum_{i\Omega_n}\sum_{\alpha'\beta'}
{\cal G}_{\alpha'\beta'}^{0}(i\omega_n-i\Omega_n)\left[\left(1+V^{A}D\right)^{-1}V^{A}-V^{A}\right]_
{a\alpha',b\beta'}(i\Omega_n).
\end{equation}
This is the diagrammatic equivalent of the self-energy approximation (~\ref{eq: affia15}) discussed in the 
previous section.

\begin{figure}[h]
\begin{center}
\epsfxsize=5in
\hspace{1cm}
\epsffile{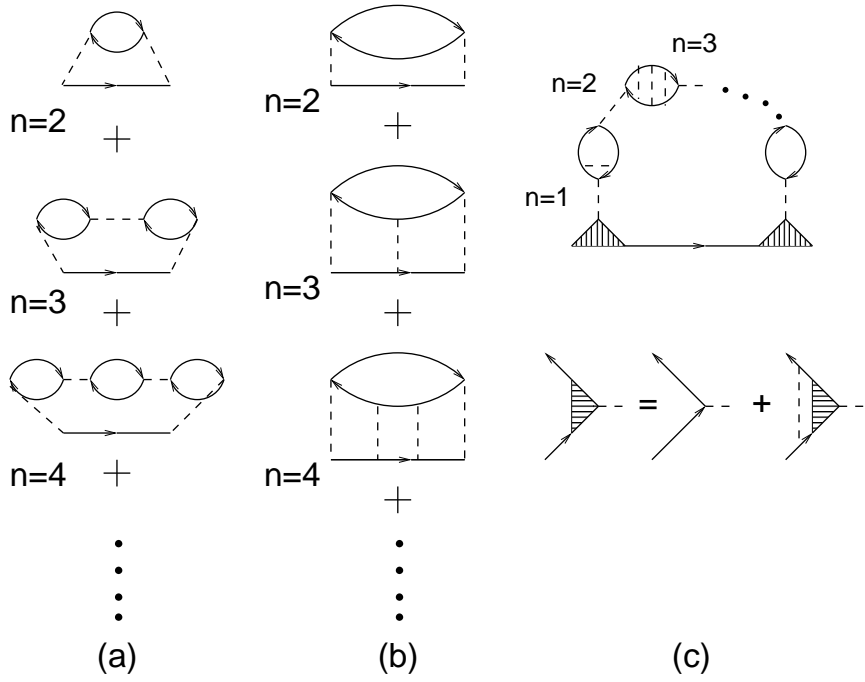}
\vspace{-2cm}
\caption{Diagrammatic content of the GRPA self-energy (~\ref{eq: affia14}). The first set of diagrams 
(a) has been traditionally used to include the effect of screening whereas the second set (b) has been 
used as the RPA self-energy in itinerant ferromagnets to include the effect of spin-waves. 
Eq.(~\ref{eq: affia14}) includes another class of diagrams, shown in (c), which captures the contribution 
of competing Hartree and exchange fluctuations. We refer to the sum of diagrams in all three sets as the 
GRPA self-energy.} 
\label{fig: allfigs}
\end{center}
\end{figure}

The diagrammatic content of this approximation is shown in Fig.~\ref{fig: allfigs}. The first set of 
diagrams on the left, (a), has been traditionally used to include the effect of screening in RPA 
self-energy~\cite{mahan}. This set of diagrams is generated by using the direct interaction as the bare 
four-point vertex in Eq.(~\ref{eq: grpa1}). The diagrams in this set can be formally summed and their 
contribution to the quasiparticle self-energy is given by 
\begin{equation}
\label{eq: grpa3}
\Sigma_{ab}(i\omega_n)=\frac{1}{\beta}\sum_{i\Omega_n}\sum_{\alpha'\beta'}
{\cal G}_{\alpha'\beta'}^{0}(i\omega_n-i\Omega_n)\left[V^{sc}-V\right]_{a\alpha,b\beta}(i\Omega_n)
\end{equation}
where we identify $V^{sc}=\left(1+VD\right)^{-1}V$ as the screened Coulomb interaction. The second set, 
(b), has been used to include the effect of spin-waves in a ferromagnet~\cite{he,kasner}. This set is 
generated by using the exchange interaction as the bare vertex. As expected, it represents only exchange 
fluctuations, which determine the spin-wave dynamics. The diagrams in this set, when formally summed, give 
\begin{equation}
\label{eq: grpa4}
\Sigma_{ab}(i\omega_n)=-\frac{1}{\beta}\sum_{i\Omega}\sum_{\alpha'\beta'}
{\cal G}_{\alpha'\beta'}^{0}(i\omega_n-i\Omega_n)\left[\left(1-V^{ex}D\right)^{-1}V^{ex}-V^{ex}\right]_
{a\alpha',b\beta'}(i\Omega_n).
\end{equation}
Eq.(~\ref{eq: grpa4}) reproduces the results for electron self-energy because of spin-waves in a single-layer 
quantum Hall system at $\nu=1$, where the physics of spin-waves is determined solely by exchange 
fluctuations~\cite{kasner}. The generalized random phase approximation introduced here includes another class 
of diagrams, (c), which represents combinations of (competing) direct and exchange fluctuations. We note that 
the set of diagrams in Fig.~\ref{fig: allfigs} arises naturally in the functional integral approach as do 
the subsets (a) and (b), when considering fluctuations around the Hartree or the exchange mean-field 
respectively. We refer to the sum of diagrams in all three sets as the GRPA self-energy. This approximation
 for the self-energy can also be derived by differentiating the grand potential with respect to the 
mean-field Green's function when the appropriate generalized random phase approximation for the grand 
potential is used~\cite{ya}. 
  

\section{Application to double layer systems}
\label{sec: adls}
In this section, we apply the formalism developed in preceding sections to a double-layer quantum Hall 
system at \emph{total} filling factor $\nu=1$. We consider a completely spin-polarized ground state so that 
the layer index and the intra-Landau level index are the only dynamical degrees of freedom. The 
properties of the Landau level wavefunctions allow us to associate a two-dimensional momentum with a 
pair of orbit-center labels~\cite{ch}. The existence of this unitary transformation between the 2D 
wavevectors and pairs of orbit-center labels permits the significant progress that can be achieved 
analytically in the following calculations.

The Hamiltonian for a bilayer system with interlayer tunneling amplitude $\Delta_t$ is 
\begin{eqnarray}
\label{eq: adls1}
\hat{H}_0-\mu\hat{N}& = &\sum_{k,\sigma}\epsilon_\sigma c^{\dagger}_{k\sigma}c_{k\sigma}, \\
\label{eq: adls2}
\hat{V} & = & \hat{V}_0 + \hat{V}_x , \\
\hat{V}_0 & = & \frac{1}{2}\sum_{k_i,\sigma_i}\langle k_1 k_2 | V_0| k_3 k_4\rangle
c^{\dagger}_{k_1\sigma_1}c^{\dagger}_{k_2\sigma_2}c_{k_4\sigma_2}c_{k_3\sigma_1}, \nonumber \\ 
\hat{V}_x & = & \frac{1}{2}\sum_{k_i,\sigma_i}\langle k_1 k_2 | V_x| k_3 k_4\rangle
c^{\dagger}_{k_1\bar{\sigma}_1}c^{\dagger}_{k_2\bar{\sigma}_2}c_{k_4\sigma_2}c_{k_3\sigma_1},
 \nonumber
\end{eqnarray}
Here, $\sigma=\uparrow=-\bar{\sigma}$ denotes a symmetric state (S) while $\sigma=\downarrow$ denotes 
the antisymmetric state (AS), and $\epsilon_\sigma=-\sigma\Delta_t/2$ are the bare single-particle energies 
measured from the chemical potential. The $k_i$ are the angular momenta which are good quantum numbers of the 
single-particle Hamiltonian in the symmetric gauge ${\bf A}=(-By/2,Bx/2,0)$; $V_0=(V_A+V_E)/2$ and 
$V_x=(V_A-V_E)/2$ are sums and differences of interlayer and intralayer Coulomb interactions. In the 
present choice of quantization axis in pseudospin-space, the interaction $V_x$ reverses pseudospins of 
the scattering particles and therefore it does not commute with the pseudospin polarization operator. 

We start with results for the Hartree-Fock mean-field approximation. The Matsubara Green's function for 
the one-body Hamiltonian $\hat{H}_0$ is
\begin{equation}
\label{eq: adls3}
{\cal G}_{k\sigma}(i\omega_n) = \frac{-1}{i\omega_n-\epsilon_\sigma}.
\end{equation}
To obtain the Hartree-Fock mean-field Green's function, we solve the Dyson equation with Hartree-Fock 
self-energy
\begin{equation}
\label{eq: adls4}
\Sigma^{HF}_\sigma=\epsilon^{0}_\sigma - \epsilon_\sigma = 
-\left[\Gamma_0(0)n_F(\epsilon^{0}_\sigma)+\Gamma_x(0)n_F(\epsilon^{0}_{\bar{\sigma}})\right],
\end{equation}
where ($\lambda=0,x,A,E$) 
\begin{equation}
\label{eq: adls5}
\Gamma_{\lambda}(\vec{q})= \frac{1}{A}\sum_{\vec{p}}V_{\lambda}(\vec{p})e^{-p^2l^2/2}
e^{i\hat{z}\cdot(\vec{q}\times\vec{p})l^2}
\end{equation}
and $n_F(x)$ is the Fermi occupation number. The $\Gamma_\lambda(\vec{q})$ are interactions between an 
electron and an exchange-hole~\cite{ch} whose center is separated by a distance $ql^2$. The Hartree-Fock 
self-energy is diagonal in the orbit center and pseudospin labels, and the Hartree-Fock mean-field 
Green's function is given by 
\begin{equation}
\label{eq: adls6}
{\cal G}^{0}_{k\sigma}(i\omega_n) = \frac{-1}{i\omega_n-\epsilon^{0}_\sigma}.
\end{equation}
At low temperatures the stable solution of Eq.(~\ref{eq: adls4}) is given by 
$\epsilon^{0}_S=-\left[\Delta_t+m_z\Gamma_E(0)\right]/2\equiv-\Delta_{SAS}/2$. Particle-hole 
symmetry at $\nu=1$ implies that $\epsilon^{0}_{AS}=-\epsilon^{0}_{S}$. The exchange enhancement of the 
symmetric-antisymmetric energy splitting, $\Delta_{sb}=m_z\Gamma_E(0)$, is determined by the interlayer 
interaction and survives in the limit of vanishing interlayer tunneling, giving rise to spontaneous phase 
coherence. The spectral function for the Hartree-Fock Green's function is given by 
$A_\sigma(\omega)=2\pi\delta(\omega-\epsilon^{0}_\sigma)$. In a mean-field approximation, the ground 
state is fully pseudospin polarized along the positive $z$-direction. The low-lying collective modes 
around this mean-field involve slow variations in the pseudospin field \emph{i. e.} the pseudospin waves. 
Fluctuations out of the $y-z$ plane in pseudospin-space correspond to the transfer of charge from one 
layer to another whereas pseudospin fluctuations in the $y-z$ plane correspond to variations in the 
relative phase between states localized in the top and the bottom layer. 

\begin{figure}[h]
\begin{center}
\epsfxsize=4in
\hspace{1cm}
\epsffile{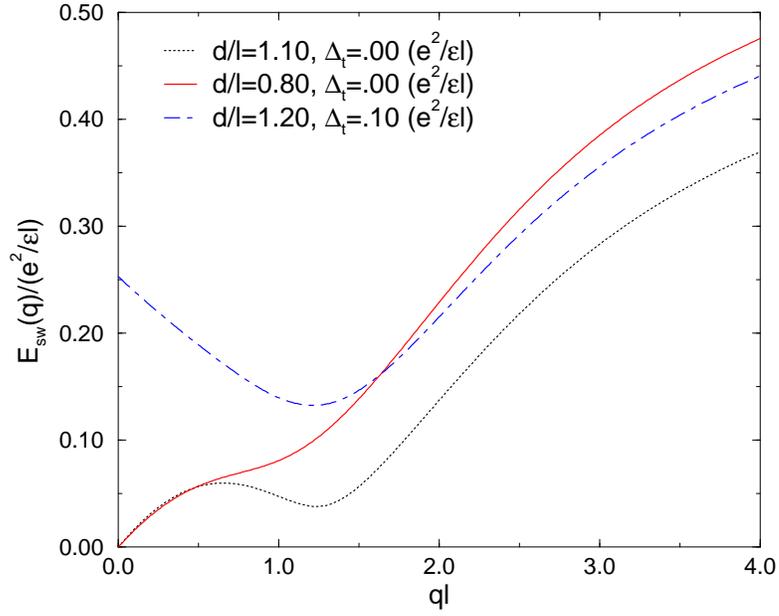}
\vspace{1cm}
\caption{Typical collective-mode dispersion for a pseudospin polarized state. For a finite interlayer 
tunneling, $\Delta_t\neq 0$, the collective mode is gapped since the U(1) symmetry in the $y-z$ plane 
is explicitly broken. The minimum in the pseudospin-wave energy near $ql\approx 1$ is because of 
\emph{competing} Hartree and exchange fluctuations present in the term $a(\vec{q})$. }
\label{fig: dispersion}
\end{center}
\end{figure}

We now evaluate the contribution of these pseudospin waves to the quasiparticle self-energy and its 
effect on the mean-field spectral function. The calculation of four point vertex $\Gamma^{(4)}$ or the 
fluctuation matrix $M$ is particularly simple in the present case. Since momentum and frequency are good 
quantum numbers, the fluctuation matrix is diagonal in these indices and effectively has only the 
pseudospin labels. In other words, the matrix equation for the four-point vertex having orbital and 
pseudospin labels, Eq.(~\ref{eq: grpa1}), reduces to an algebraic equation with matrices having only 
pseudospin labels. Furthermore, because of the constraints which ensure that the Hartree-Fock susceptibility 
$D$ is nonzero, only a $2\times 2$ submatrix of the entire $4\times 4$ matrix $V^{A}D$ is nonzero. This 
non-vanishing $2\times 2$ submatrix is given by  
\begin{equation}
\label{eq: adls7}
\langle\sigma_1\sigma_2|V^{A}(\vec{q})D(i\Omega_n)|\sigma_3\sigma_4\rangle = 
\left[\begin{array}{cc}
(v_x-\Gamma_0)\frac{1}{i\Omega_n+\Delta_{SAS}} & -(v_x-\Gamma_x)\frac{1}{i\Omega_n-\Delta_{SAS}} \\ 
(v_x-\Gamma_x)\frac{1}{i\Omega_n+\Delta_{SAS}} & -(v_x-\Gamma_0)\frac{1}{i\Omega_n-\Delta_{SAS}}
\end{array} \right].
\end{equation} 
Here $2\pi l^2 v_x(\vec{q})=e^{-q^2l^2/2}V_x(\vec{q})$ represents the effect 
of electrostatic fluctuations whereas $\Gamma(\vec{q})$ represent the effect of exchange fluctuations. 
The entire $4\times 4$ fluctuation matrix is a trivial extension of Eq.(~\ref{eq: adls7}). The roots of 
the equation $\det\left[1+V^{A}(\vec{q})D(i\Omega)\right]=0$ give the bilayer pseudospin-wave 
dispersion~\cite{haf,apb} $i\Omega=E_{sw}(\vec{q})$ where 
\begin{eqnarray}
\label{eq: adls8}
E_{sw}^2(\vec{q}) = & a(\vec{q})\cdot b(\vec{q}) & = 
\left[\Delta_{SAS}-\Gamma_A(\vec{q})+2v_x(\vec{q})\right]\cdot
\left[\Delta_{SAS}-\Gamma_E(\vec{q})\right].
\end{eqnarray}
Thus, the functional integral approach reproduces results obtained earlier by diagrammatics~\cite{haf} 
and by the single-mode approximation~\cite{apb}. Figure~\ref{fig: dispersion} shows typical pseudospin-wave 
dispersions. For a finite interlayer tunneling, $\Delta_t\neq 0$, the collective mode is gapped at zero 
wavevector because the U(1) symmetry in the $y-z$ plane in pseudospin-space is explicitly broken. As $d$ 
increases, the minimum in the pseudospin-wave spectrum near $ql\approx 1$ reaches zero at a critical 
layer separation $d_{cr}$. In Hartree-Fock theory, at this point, the ground state changes from a 
uniformly pseudospin-polarized state to a pseudospin-density wave state~\cite{cote}. This softening of 
pseudospin-wave at a finite wavevector has been associated with the phase transition~\cite{apb,sqm} from
 a phase-coherent quantum Hall state to a compressible state.

Using the mean-field Green's function (~\ref{eq: adls6}) and the $4\times 4$ fluctuation matrix we 
arrive at the following analytical expression for zero-temperature symmetric state 
self-energy
\begin{equation}
\label{eq: adls9}
\Sigma_{S}(i\omega_n)= \frac{2\pi l^2}{A}\sum_{\vec{p}}
\frac{(E_{sw}(\vec{p})+\Delta_{SAS})^2}{2E_{sw}(\vec{p})}
\left[\frac{\epsilon(\vec{p})-E_{sw}(\vec{p})}{i\omega_n-E_{sw}(\vec{p})-\epsilon^0_{AS}}\right] 
= -\Sigma_{AS}(-i\omega_n),
\end{equation}
where $\epsilon(\vec{p})=\left[a(\vec{p})+b(\vec{p})\right]/2$. This remarkably simple expression is the 
first principal result of this work. The Dyson equation relating the full Green's function to the 
mean-field Green's function and the fluctuation self-energy is 
\begin{equation}
\label{eq: adls10}
\left[{\cal G}_{\sigma}(i\omega_n)\right]^{-1}=\left[{\cal G}^{0}_{\sigma}(i\omega_n)\right]^{-1}
+\Sigma_{\sigma}(i\omega_n).
\end{equation}
We obtain the quasiparticle spectral function from the retarded GRPA self-energy by analytically 
continuing (~\ref{eq: adls9}) to real frequencies, $i\omega_n\rightarrow\omega+i\eta$, 
\begin{equation}
\label{eq: adls11}
A_S(\omega)=-2 \mbox{ Im } G_S(\omega+i\eta)= 
-2 \mbox{ Im }\frac{1}{\omega+i\eta-\epsilon^{0}_S-\Sigma_S(\omega+i\eta)} = A_{AS}(-\omega).
\end{equation}

\begin{figure}[h]
\begin{center}
\epsfxsize=4in
\hspace{1cm}
\epsffile{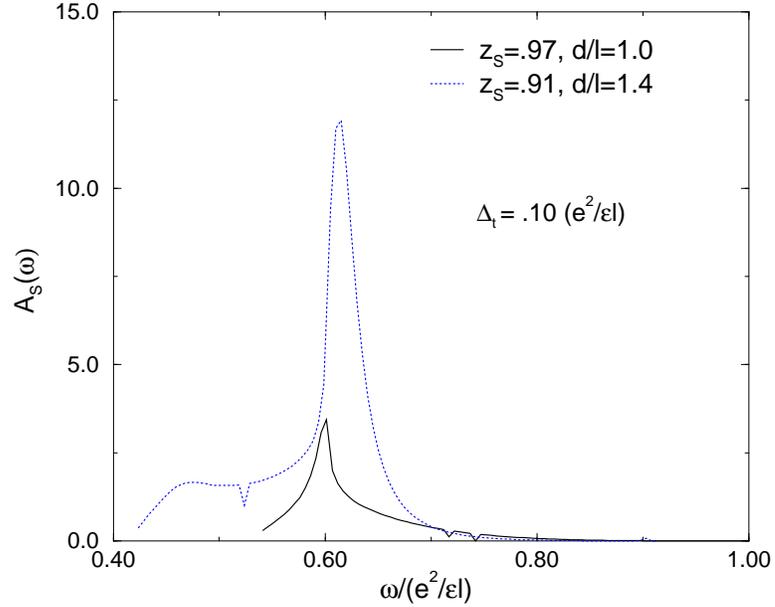}
\vspace{1cm}
\caption{ Continuum part of the symmetric-state spectral function: The zero of energy is at the chemical
 potential. As the interlayer spacing increases, the spectral weight at positive frequencies increases 
and the spectral weight $z_S$ in the $\delta$-function peak at the symmetric-state quasiparticle pole 
decreases. This presence of spectral weight at positive energies has been detected by 
Manfra {\it et al.}~\cite{manfra}}
\label{fig: spectral}
\end{center}
\end{figure}

\begin{figure}[h]
\begin{center} 
\epsfxsize=4in
\hspace{1cm}
\epsffile{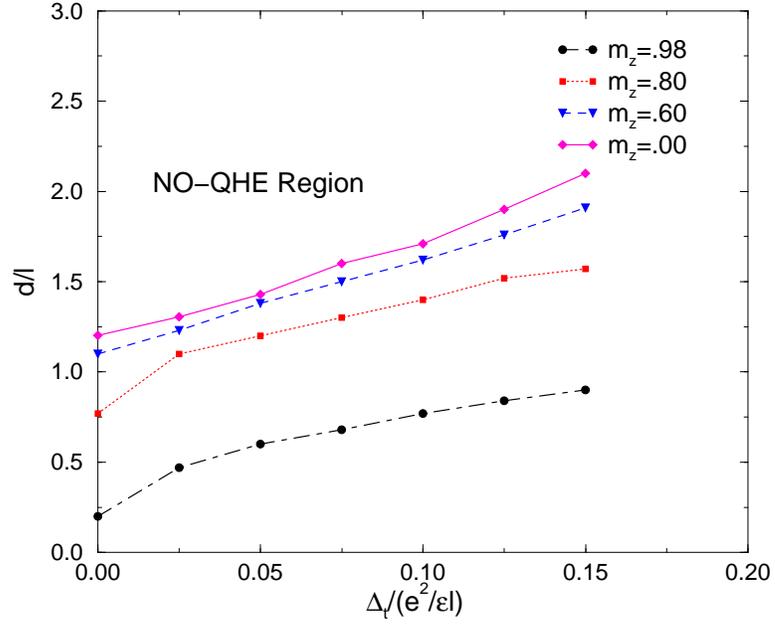}
\vspace{1cm}
\caption{Renormalized polarization contours. The rapid suppression of the order parameter \emph{only} 
close to the phase-boundary indicates that interactions between the collective modes can be ignored 
for most part of the parameter space in the phase-coherent regime.}
\label{fig: phasediagram}
\end{center}
\end{figure}

\begin{figure}[b]
\begin{center}
\epsfxsize=4in
\hspace{1cm}
\epsffile{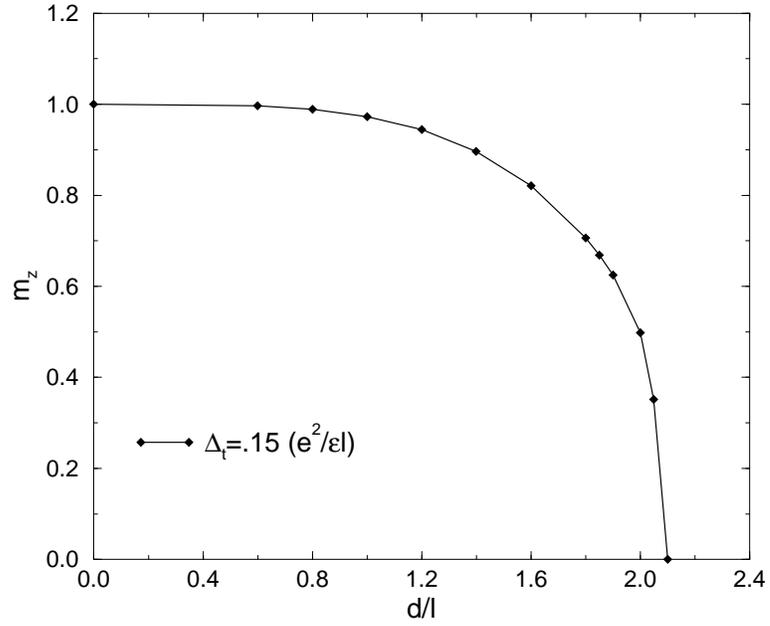}
\vspace{1cm}
\caption{Typical renormalized polarization at $T=0$. The mean-field polarization is independent of the 
changes in the nature of collective mode spectrum which occur as $d$ approaches the critical layer 
separation, $d_{cr}/l=2.15$ for $\Delta_t=.15 (e^2/\epsilon l)$. The renormalized polarization vanishes 
rapidly close to the phase-boundary because of increasing quantum fluctuations.}
\label{fig: polarization}
\end{center}
\end{figure}

This spectral function has a $\delta$-function contribution at frequency $\omega^{*}$ which satisfies 
the Dyson equation $\omega^{*}-\epsilon^0_S=\Sigma_S(\omega^{*}+i\eta)$. The continuum piece of 
$A_S(\omega)$ is nonzero in the region where the retarded self-energy $\Sigma_S(\omega+i\eta)$ has a 
branch cut, \emph{i. e.}, at frequencies in the interval $I_S\equiv\left[\omega_{<S},\omega_{>S}\right]=
\left[\epsilon_{AS}^{0}+E_{sw}^{min},\epsilon^{0}_{AS}+\Delta_{SAS}\right]$. Here $E_{sw}^{min}$ denotes 
the minimum of the pseudospin-wave energy near $ql\approx 1$ and 
$\Delta_{SAS}=E_{sw}(q\rightarrow\infty)$ is the mean-field splitting. Explicitly, we can write the 
spectral function as 
\begin{equation}
\label{eq: adls12}
A_S(\omega)= 2\pi z_S\delta(\omega-\omega^{*})+\theta(\omega-\omega_<)\theta(\omega_>-\omega)
\frac{-2\mbox{ Im}\Sigma_S(\omega)}{\left[\omega-\epsilon^{0}_S-\mbox{Re}\Sigma_S(\omega)\right]^2 +
\left[\mbox{Im}\Sigma_S(\omega)\right]^2},
\end{equation}
where $z_S=(1-\partial\Sigma_S/\partial\omega|_{\omega^{*}})^{-1}<1$ is the quasiparticle 
renormalization factor. The mean-field spectral function $A_S(\omega)=2\pi\delta(\omega-\epsilon^{0}_S)$
 is recovered by neglecting the fluctuation self-energy, which implies  
$\omega^{*}=\epsilon^{0}_S$ and $z_S=1=z_{AS}$. When we include the fluctuation self-energy, the 
spectral weight at the $\delta$-function peak is reduced to $z_S<1$ and is distributed into a continuum 
piece at positive energies, $\omega>\mu$. Figure~\ref{fig: spectral} shows the evolution of 
\emph{continuum} part of the symmetric-state spectral function with increasing layer separation. We see 
that the spectral weight at positive energies, an indicator of the strength of quantum fluctuations in 
the pseudospin polarized mean-field ground state, increases with increasing layer separation and as required 
by the sum-rule, the quasiparticle renormalization factor $z_S$ decreases. The presence of the spectral 
weight at positive energies has been detected at low temperatures by Manfra {\it et al.} in optical 
absorption experiments~\cite{manfra}. The redistribution of the spectral weight indicates the mixing of 
many-body states having antisymmetric quasiparticles into the completely pseudospin-polarized ground state. 
One measure of this mixing is the suppression of the order parameter 
\begin{equation}
\label{eq: adls13}
m_z=(n_{S}-n_{AS})=\int^{\infty}_{-\infty}\frac{d\omega}{2\pi}n_F(\omega)
\left[A_S(\omega)-A_{AS}(\omega)\right]
\end{equation}
from its mean-field value $m_z=1$. Using the particle-hole symmetry at $\nu=1$ we get 
$A_S(\omega)=A_{AS}(-\omega)$. This leads to the following simple expression for the zero-temperature 
renormalized order parameter
\begin{eqnarray}
\label{eq: adls14}
m_z(T=0) & = & 2z_S-1, \\
\label{eq: adls15}
z_S & = & 
\left[1+\frac{2\pi l^2}{A}\sum_{\vec{p}}\frac{(E_{sw}(\vec{p})+\Delta_{SAS})^2}{2E_{sw}(\vec{p})}
\frac{\epsilon(\vec{p})-E_{sw}(\vec{p})}{(\omega^{*}-E_{sw}(\vec{p})-\epsilon_{AS}^{0})^2}\right]^{-1}.
\end{eqnarray}
Eq.(~\ref{eq: adls15}) is the second principal result of this work. Figure~\ref{fig: phasediagram} shows 
contours of renormalized pseudospin polarization in the parameter space $(d,\Delta_t)$. In the 
mean-field approximation, the pseudospin polarization is not susceptible to the softening of the collective 
mode at a finite wavevector as $d\rightarrow d_{cr}$. When we include the effect of collective modes, the 
order parameter is strongly suppressed as the layer separation approaches the critical
 layer separation. We plot the typical dependence of renormalized polarization $m_z$ on interlayer 
spacing $d$ in Fig.~\ref{fig: polarization}. We interpret the strong suppression of the order parameter 
because of the quantum fluctuations as a precursor of the phase-transition to a compressible 
state~\cite{apb,sqm}.


\section{Adiabatic Approximation}
\label{sec: aa}
In the preceding sections, we presented the generalized random phase approximation for the quasiparticle 
self-energy using a functional integral approach and discussed its equivalent approximation in diagrammatic
 perturbation theory. In this section, we present an adiabatic approximation for the 
fluctuation self-energy~\cite{atk}. This approximation is intuitively appealing and is applicable to a 
large class of problems, including metallic ferromagnets, where the effects of dispersive band structure
 are important~\cite{atk}. We will show that for bilayer systems, the fluctuation self-energy 
calculated using the adiabatic approximation is identical to the GRPA self-energy. 

The starting point for this approximation is the Hartree-Fock mean-field-theory Hamiltonian where we 
assume that the ground state is pseudospin polarized. The mean-field Hamiltonian couples the pseudospin 
polarization $m_z$ to the self-consistent field $\phi^{0}$ generated by other electrons. In the 
adiabatic approximation, we treat this field $\phi$ as a \emph{dynamical} variable and consider small 
fluctuations of this field from its self-consistent value $\phi^{0}$. We expand the Hamiltonian, 
considered as a function of quasiparticles \emph{and} the field, around the self-consistent value of the
 field. The leading-order term in this expansion gives an effective interaction between the mean-field 
quasiparticles and the collective modes around that mean-field. We approximate the fluctuation 
self-energy by a single collective-mode exchange diagram similar to the approximation commonly used to 
treat phonon-exchange in metals~\cite{mahan}. 

It is particularly easy to evaluate this self-energy for a bilayer quantum Hall system because the 
Landau level bands have no dispersion and because particle-hole pair-momentum is a good quantum 
number~\cite{ch}. We start with the Hartree-Fock Hamiltonian in the pseudospin polarized state
\begin{equation}
\label{eq: aa1}
\hat{H}_{HF}=-\frac{\Delta_t}{2}\sum_{k,\sigma}\sigma c^{\dagger}_{k\sigma}c_{k\sigma}+
\sum_{k'\sigma'k\sigma}\langle k'\sigma' | U_{HF}|k\sigma\rangle 
c^{\dagger}_{k'\sigma'}c_{k\sigma},
\end{equation}
where the mean-field potential $\hat{U}_{HF}(\vec{q})$ is given by
\begin{equation}
\label{eq: aa2}
\hat{U}_{HF}(\vec{q}) =  
\left[ \tau^{0}\cdot\left(v_0^{\vec{q}}-\frac{\Gamma_A^{\vec{q}}}{2}\right)\langle m^{0}_{-\vec{q}}\rangle+
\tau^{x}\cdot\left(v_x^{\vec{q}}-\frac{\Gamma_A^{\vec{q}}}{2}\right)\langle m^{x}_{-\vec{q}}\rangle-
\tau^{y}\cdot\frac{\Gamma_E^{\vec{q}}}{2}\langle m^{y}_{-\vec{q}}\rangle-
\tau^{z}\cdot\frac{\Gamma_E^{\vec{q}}}{2}\langle m^{z}_{-\vec{q}}\rangle \right].
\end{equation}
Here $\tau^{\mu}$ are the Pauli matrices acting in the pseudospin-space, $v_\lambda^{\vec{q}}$ and 
$\Gamma_{\lambda}^{\vec{q}}$ are the direct and exchange interactions, and 
\begin{equation}
\label{eq: aa3}
m^{\mu}_{\vec{q}}=\sum_{k',k}\sum_{\sigma'\sigma}\langle k'| e^{-i\vec{q}\cdot\hat{r}}|k\rangle 
c^{\dagger}_{k'\sigma'}\tau^{\mu}_{\sigma'\sigma}c_{k\sigma}
\end{equation}
are the charge and pseudospin densities. To obtain the dynamics of the pseudospin waves and their 
effective interaction with the mean-field quasiparticles, we linearize Eq.(~\ref{eq: aa2}) around the 
pseudospin polarized state given by $\langle m^{z}_{\vec{q}}\rangle=A\delta_{q,0}/2\pi l^2$. The transverse 
fluctuations $m^{x}_{\vec{q}}$ correspond to charge-imbalance modulations with wavevector $\vec{q}$ 
whereas the fluctuations in $m^{y}_{\vec{q}}$ represent the variation of relative phase between states 
localized in the top and the bottom layer. These fluctuations are canonically conjugate, 
$[m^{x},m^{y}]=2im^{z}\approx 2iA/2\pi l^2$, and therefore we quantize them using bosonic creation and 
annihilation operators
\begin{eqnarray}
\label{eq: aa4} 
m^{y}_{\vec{q}}=\sqrt{\frac{A}{2\pi l^2}}\left(a_{\vec{q}}+a^{\dagger}_{-\vec{q}}\right), 
& \mbox{\hspace{.5cm} } & 
m^{x}_{\vec{q}}=i\sqrt{\frac{A}{2\pi l^2}}\left(a_{\vec{q}}-a^{\dagger}_{-\vec{q}}\right).
\end{eqnarray}
The Hamiltonian for the bosonic modes introduced in (~\ref{eq: aa4}) is obtained from the Hartree-Fock 
energy functional 
\begin{equation}
\label{eq: aa5}
E_{HF}\left[\vec{m}\right]=
-\frac{\Delta_t}{2}m^{z}_{\vec{q}=0}+
\frac{2\pi l^2}{A}\sum_{\vec{p}}
\left[
\left(v_x^{\vec{p}}-\frac{\Gamma_E^{\vec{p}}}{2}\right)m^{x}_{\vec{p}}m^{x}_{-\vec{p}} -
\frac{\Gamma_E^{\vec{p}}}{2}
\left(m^{y}_{\vec{p}}m^{y}_{-\vec{p}}+m^{z}_{\vec{p}}m^{z}_{-\vec{p}}\right)
\right].
\end{equation}
We expand energy functional (~\ref{eq: aa5}) around the mean-field state 
$\langle m^{z}_{\vec{q}}\rangle=A\delta_{q,0}/2\pi l^2$ and quantize the fluctuations using 
Eq.(~\ref{eq: aa4}). A simple calculation gives the following Hamiltonian for the pseudospin-waves
\begin{equation}
\label{eq: aa6}
\hat{H}^{sw}=\sum_{\vec{p}}
\left[\epsilon_{\vec{p}}a^{\dagger}_{\vec{p}}a_{\vec{p}}+\frac{\lambda_{\vec{p}}}{2}
\left(a^{\dagger}_{\vec{p}}a^{\dagger}_{-\vec{p}} + a_{\vec{p}}a_{-\vec{p}}\right)
\right],
\end{equation}
where $\epsilon(\vec{p})=[\Delta_{SAS}+v_x-\Gamma_0]=[a(\vec{p})+b(\vec{p})]/2$ and 
$\lambda(\vec{p})=[v_x-\Gamma_x]=[a(\vec{p})-b(\vec{p})]/2$. The anisotropy of Coulomb interaction in 
the pseudospin-space is reflected in the fact that the number operator 
$\hat{n}_{\vec{p}}=a^{\dagger}_{\vec{p}}a_{\vec{p}}$ does not commute with the Hamiltonian 
(~\ref{eq: aa6}). The Hamiltonian is diagonalized by Bogoliubov transformation~\cite{bg} and we get 
non-interacting pseudospin-waves with the dispersion 
$E_{sw}^2(\vec{q})=\epsilon^2(\vec{q})-\lambda^2(\vec{q})=a(\vec{q})\cdot b(\vec{q})$ obtained in 
preceding sections. The effective interaction between pseudospin waves and the mean-field quasiparticles is 
obtained by expanding the Hartree-Fock Hamiltonian (~\ref{eq: aa1}) in terms of pseudospin-wave 
operators. At leading order we get
\begin{eqnarray}
\label{eq: aa7}
\hat{H}^{e-sw}_{k'k}& = &\sum_{\vec{p}}\frac{1}{\sqrt{2\pi l^2 A}}
\left(c^{\dagger}_{k'S}c_{kAS}\cdot a^{\dagger}_{-\vec{p}}-
c^{\dagger}_{k'AS}c_{kS}\cdot a_{\vec{p}}\right) M_{k'k}(\vec{p}) \nonumber \\
& - & \sum_{\vec{p}}\frac{1}{\sqrt{2\pi l^2 A}}
\left(c^{\dagger}_{k'S}c_{kAS}\cdot a_{\vec{p}}-
c^{\dagger}_{k'AS}c_{kS}\cdot a^{\dagger}_{-\vec{p}}\right) N_{k'k}(\vec{p}),
\end{eqnarray}
where the interaction matrix elements are given by
\begin{eqnarray}
\label{eq: aa8} 
M_{k'k}(\vec{q})=i(2\pi l^2)\langle k | e^{-i\vec{q}\cdot\hat{r}}|k'\rangle 
\left[v_x(\vec{q})-\Gamma_0(\vec{q})\right], \\
\label{eq: aa9} 
N_{k'k}(\vec{q})=i(2\pi l^2)\langle k | e^{-i\vec{q}\cdot\hat{r}}|k'\rangle 
\left[v_x(\vec{q})-\Gamma_x(\vec{q})\right].
\end{eqnarray}

A straightforward albeit lengthy calculation that follows line similar to the earlier 
work~\cite{mahan,kasner} gives fluctuation self-energy expression (~\ref{eq: adls9}) identical to the 
one obtained using the functional integral approach or the diagrammatic approximation. 


\section{Summary and Discussion}
\label{sec: discussion}
This paper discusses an approximation for the self-energy of interacting electrons that goes beyond the 
venerable random phase approximation (RPA). Depending on context the RPA describes in an approximate 
way the interactions of electrons with fluctuations in either direct or exchange particle-hole channels. 
There are many circumstances, however, where both channels must be treated on an equal footing if the 
qualitative physics is to be captured. The objective of the approximation scheme discussed here is to fill 
this need. We have developed our approximation in the language of a path-integral approach developed by 
Kerman {\it et al.}, that, unlike standard auxiliary-field path integral approaches, enables expansions to 
be made around Hartree-Fock mean-field states in which the quasiparticles experience both direct and 
exchange fluctuating potentials. The same approximation for the self-energy can be derived by 
differentiating the approximation for the grand potential that follows from this generalized RPA scheme, 
with respect to the mean-field Green's function. We have identified the infinite subset of many-body 
perturbation theory Feynman diagrams that are included in this self-energy approximation. As far as we are 
aware, this self-energy approximation has not been considered previously. 

We have applied our formalism to the ground state of bilayer quantum Hall systems at filling factor 
$\nu=1$. This ground state has a broken symmetry in which phase-coherence is spontaneously established 
between the two layers in order to take advantage of interlayer exchange and improve interlayer 
correlations. Its collective excitations involve both fluctuations in the interlayer phase and 
fluctuations in the difference between charge densities in each layer. In order to describe the energetics of 
these fluctuations, it is necessary to include both direct and exchange particle-hole interactions. 
Bilayer quantum Hall systems therefore provide an example of a situation where the standard RPA will fail 
quite badly. Our approximate self-energy properly describes the interactions between quasiparticles in 
this system and the collective excitations of its order parameter field. Bilayer quantum Hall systems 
provide an interesting example for our approach because the importance of fluctuations for the ground 
state properties can be tuned between zero and large values by varying the layer separation. It is now 
well established by both theory and experiment that bilayer quantum Hall systems have a quantum phase 
transition~\cite{apb,sqm,sgm} at a finite value of $d$ where the phase-coherence is lost. Our 
approximate self-energy calculation correctly obtains an order parameter that vanishes beyond a critical 
value of $d$, however, we do not expect it to be reliable very close to the transition point, whether it is 
first order or continuous. The bilayer quantum Hall case is also favorable because particle-hole excitations 
of the ground state have momentum as a good quantum number and there is only one excitation at each 
momentum. In effect, there is only one excited state at each momentum, the collective excitation. Much 
as in the case of a one-dimensional electron system, there is no particle-hole continuum in the excitation 
spectrum. This greatly simplifies the calculation and allows us to derive a remarkably simple analytic 
expression for the self-energy. 

We have confirmed the physical content of our theory by demonstrating its equivalence, for the case of 
bilayer quantum Hall systems, to an adiabatic approximation in which the Hartree-Fock single-particle 
Hamiltonian is expressed in terms of the order-parameter field, via the density matrix. 
The possibility of specifying the density matrix that appears in the Hartree-Fock single-particle Hamiltonian 
in terms of the order parameter field is unique to quantum Hall systems. Therefore this simple relationship 
cannot 
always be established. Nevertheless, this example demonstrates the ability of the approximation we discuss to 
accurately capture the effect of important fluctuations on quasiparticle properties. 

We believe that our approximate self-energy will be useful for other situations as well. A potentially 
interesting example is provided by the case of superconductors for which coupled fluctuations in the 
phase of the order parameter and of the charge density form elementary excitations and both pairing and 
electrostatic energies are important. In the case of superconductors it is particle-particle channel 
fluctuations are important for phase-fluctuations, unlike the case of bilayer quantum Hall systems. The 
problems can be made quite similar, however, by performing a particle-hole transformation for, say, the 
down-spins of the superconductor, turning the superconducting order into easy-plane ferromagnetism. In fact, 
going in the opposite direction, the order of a bilayer quantum Hall ferromagnet can be regarded as an 
electron-hole pair condensate by making a particle-hole transformation in one of the layers. In the case 
of a superconductor it is well-known that the long-range of the electron-electron interaction turns the 
collective modes into gapped plasmons, usually leading to high accuracy for the mean-field treatment of a 
superconductor. This can change, however, if the collective excitations become soft at large wavevectors 
as they do in bilayer quantum Hall systems. Such softness could be associated, for example, with partial 
Fermi surface nesting. 

In this work, we have concentrated on \emph{quantum} fluctuation effects at \emph{zero} temperature. 
It is easy to obtain corresponding expressions for finite temperature self-energy, though they are 
somewhat cumbersome. In the limit of vanishing layer separation, $d\rightarrow 0$, the finite-temperature 
expressions reproduce, with the appropriate identification of pseudospin with spin, the spin-wave 
contribution to the electron self-energy in a single-layer system~\cite{kasner}. We have neglected the 
form-factors which encode the effects of finite well-widths since their inclusion does not change the results 
qualitatively. They can be incorporated without difficulty when modeling a specific system. 

\section{Acknowledgements}
This work was supported in part by the Robert A. Welch Foundation, by the Indiana 21st Century Fund, and 
by the NSF under grant DMR0115947.


\begin{thebibliography}{99}
\bibitem{kmoon} K. Moon {\it et al.}, Phys. Rev. B {\bfseries 51}, 5138 (1995).
\bibitem{ky} K. Yang {\it et al.}, Phys. Rev. B {\bfseries 54}, 11644 (1996).
\bibitem{nayak} The full phase diagram of bilayer quantum Hall system near filling factor $\nu=1$ is 
still under active investigations. See for example, J. I. Watanabe and T. Nakajima, Phys. Rev. B 
{\bfseries 62} 12997 (2000); 
E. Demler, C. Nayak, and S. Das Sarma, Phys. Rev. Lett. {\bfseries 86}, 1853 (2001); 
L. Radzihovsky, cond-mat/0104128.
\bibitem{kyang} K. Yang, Phys. Rev. Lett. {\bfseries 87}, 056802 (2001).
\bibitem{qhereviews} For reviews of bilayer QHE systems, see S. M. Girvin and A. H. MacDonald in 
{\it Perspectives in Quantum Hall Effects}, edited by S. Das Sarma and Aron Pinczuk; 
S. Das Sarma and E. Demler, Solid State Commun. {\bfseries 117}, 141 (2001); 
S. M. Girvin, {\it Physics Today}, pp. 39-45 June (2000). 
\bibitem{haf} H. A. Fertig, Phys. Rev. B {\bfseries 40}, 1087 (1989).
\bibitem{apb} A. H. MacDonald, P. M. Platzman, and G. S. Boebinger, Phys. Rev. Lett. 
{\bfseries 65}, 775 (1990).
\bibitem{brey} L. Brey, Phys. Rev. Lett. {\bfseries 65}, 903 (1990).
\bibitem{wenzee} X. G. Wen and A. Zee, Phys. Rev. Lett. {\bfseries 69}, 1811 (1992).
\bibitem{ezawa} Z. F. Ezawa and A. Iwazaki, Int. J. Mod. Phys. B {\bfseries 19} 3205 (1992).
\bibitem{sqm} S. Q. Murphy, J. P. Eisenstein, G. S. Boebinger, L. N. Pfeiffer, and K. W. West, 
Phys. Rev. Lett. {\bfseries 72}, 728 (1994).
\bibitem{az} A. H. MacDonald and S. C. Zhang, Phys. Rev. B {\bfseries 49}, 17208 (1994).
\bibitem{sgm} J. Schliemann, S. M. Girvin, and A. H. MacDonald, Phys. Rev. Lett. {\bfseries 86}, 
1849 (2001).
\bibitem{leon} L. Balents and L. Radzihovsky, Phys. Rev. Lett. {\bfseries 86}, 1825 (2001).
\bibitem{ady} A. Stern, S. M. Girvin, A. H. MacDonald, and N. Ma, Phys. Rev. Lett. {\bfseries 86}, 
1829 (2001).
\bibitem{fogler} M. M. Fogler and F. Wilczek, Phys. Rev. Lett. {\bfseries 86}, 1833 (2001).
\bibitem{ynjahm} Y. N. Joglekar and A. H. MacDonald, Physica E {\bfseries 6}, 627 (2000); cond-mat/9909057.
\bibitem{mahan} See, for example, G. D. Mahan, {\it Many-Particle Physics} 
(Plenum Press, New York, 1981). 
\bibitem{he} J. A. Hertz and D. M. Edwards, J. Phys. F: Met. Phys. {\bfseries 3}, 2174 (1973); 
{\bfseries 3}, 2191 (1973). 
\bibitem{kasner} M. Kasner and A. H. MacDonald, Phys. Rev. Lett. {\bfseries 76}, 3204 (1996); 
M. Kasner, J. J. Palacios, and A. H. MacDonald, Phys. Rev. B {\bfseries 62}, 2640 (2000).
\bibitem{negele} J. W. Negele and H. Orland, {\it Quantum Many-Particle Systems} (Addison-Wesley, 
New York, 1988).
\bibitem{kerman} A. K. Kerman, S. Levit, and T. Troudet, Ann. Phys. {\bfseries 148}, 436 (1983).
\bibitem{ya} Y. N. Joglekar and A. H. MacDonald, cond-mat/0105620.
\bibitem{ch} C. Kallin and B. I. Halperin, Phys. Rev. B {\bfseries 30}, 5655 (1984).
\bibitem{cote} R. C\^ot\'e, L. Brey, and A. H. MacDonald, Phys. Rev. B {\bfseries 46}, 10239 (1992).
\bibitem{manfra} M. J. Manfra {\it et al.}, Physica E {\bfseries 6}, 590 (2000); cond-mat/9809373.
\bibitem{atk} A. H. MacDonald, T. Jungwirth, and M. Kasner, Phys. Rev. Lett. {\bfseries 81}, 705 (1998).
\bibitem{bg} See, for example, J. R. Schrieffer, {\it Theory of Superconductivity}, Perseus Books, 
New York, 1999); 
N. N. Bogoliubov, Nuovo Cimento {\bfseries 7}, 794 (1958).
\end{thebibliography}
\end{document}